**Activity-Brightness Correlations for the Sun and Sun-like Stars.**

D. G. Preminger, G. A. Chapman and A. M. Cookson

San Fernando Observatory, Department of Physics and Astronomy, California State University Northridge, Northridge, California, USA

**Abstract**

We analyze the effect of solar features on the variability of the solar irradiance in three different spectral ranges. Our study is based on two solar-cycles' worth of full-disk photometric images from the San Fernando Observatory, obtained with red, blue and Ca II K-line filters. For each image we measure the photometric sum, $\Sigma$, which is the relative contribution of solar features to the disk-integrated intensity of the image. The photometric sums in the red and blue continuum, $\Sigma_r$ and $\Sigma_b$, exhibit similar temporal patterns: they are negatively correlated with solar activity, with strong short-term variability and weak solar-cycle variability. However, the Ca II K-line photometric sum, $\Sigma_K$, is positively correlated with solar activity and has strong variations on solar-cycle timescales. We show that we can model the variability of the Sun's bolometric flux as a linear combination of $\Sigma_r$ and $\Sigma_K$. We infer that, over solar-cycle timescales, the variability of the Sun's bolometric irradiance is directly correlated with spectral line variability, but inversely correlated with continuum variability. Our blue and red continuum filters are quite similar to the Strömgren b and y filters used to measure stellar photometric variability. We conclude that active stars whose visible continuum brightness varies inversely with activity, as measured by the Ca HK index, are displaying a pattern that is similar to that of the Sun, i.e. radiative variability in the visible continuum that is spot-dominated.

## 1. Introduction

Total Solar Irradiance [TSI] is the bolometric solar energy flux received at the Earth at its average distance of 1 AU, measured since 1976 by a succession of spacecraft instruments. Figure 1 shows the composite TSI record, compiled by the PMOD World Radiation Center (Fröhlich 2000, 2006). Sunspot cycles 21, 22, and 23 are clearly visible. Sunspots are manifestations of magnetic active regions, and their properties can be measured directly on solar images and magnetograms. Solar activity modulates TSI in a complex way: at solar maximum, when there are a large number of active regions, TSI is high on average, but TSI drops sharply every time a large sunspot group transits the solar disk.

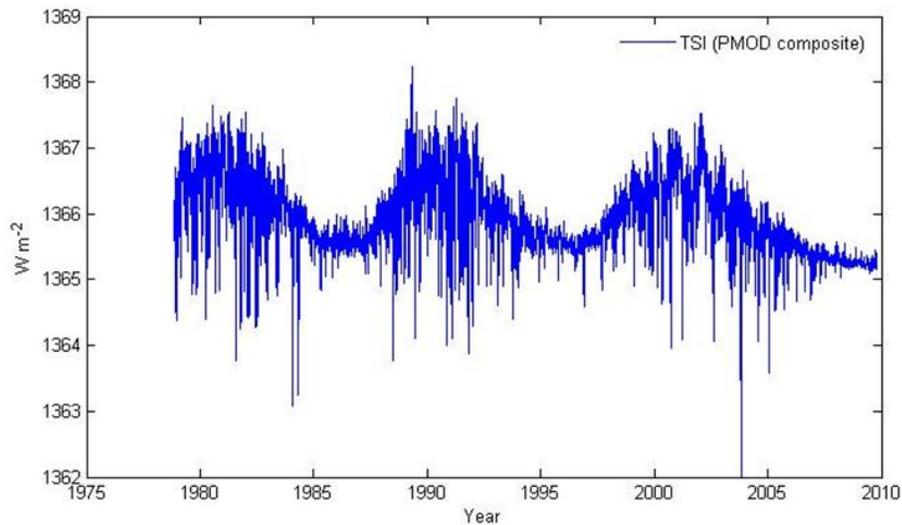

Figure 1. Total Solar Irradiance (PMOD Composite). TSI is a bolometric measurement.

Several Sun-like stars also show evidence of magnetic activity, and their variability has been studied extensively with photometry and spectroscopy (Radick et al. 1998; Lockwood et al. 2007; Hall et al. 2009). For stars, magnetic activity cannot be observed directly; instead, Ca HK emission index is taken as a proxy for magnetic activity, and Strömgren b+y brightness is taken as a measure of bolometric brightness. One of the fundamental questions asked regarding these solar analogs is: do brightness and activity for these stars vary in a Sun-like way? Observations by Lockwood et al (2007) and Hall (2009) have shown that Sun-like stars seem to fall into two groups, according to the correlation between their b+y brightness and Ca HK emission : highly active stars display an inverse correlation, while less active stars a direct one. They concluded that there is some critical stellar activity level at which the activity-brightness relation reverses. The Sun was put into the less active group because we observe a direct correlation between time-averaged TSI and solar activity level.

In this work we examine the relationship between TSI and photometric solar data from the San Fernando Observatory and discuss what it reveals about the activity-brightness correlations for the Sun-as-a-star. We then compare our results with the activity-brightness relations that have been measured for Sun-like stars.

## 2. Solar Photometry at the San Fernando Observatory

### 2.1 Observing program

At the San Fernando Observatory [SFO], full-disk photometric solar images are recorded daily with two purpose-built telescopes, CFDT1 (5" pixels) and CFDT2 (2.5" pixels), described in Chapman et al. (1997). The SFO observational record now includes about two solar-cycles' worth of solar images at different wavelengths. The images pertinent to this study are those at 672.3 nm with a 10-nm bandpass (red), 472.3 nm with a 10-nm bandpass (blue), and 393.4 nm with a 1-nm bandpass (Ca II K). Using the Kurucz solar flux atlas (Kurucz 2005), we determined that photospheric spectral lines block ~2% and ~11% of the continuum flux in the passbands of the red and blue filters, respectively. Thus our red and blue images show primarily continuum emission, from the photosphere.  Our Ca II K filter spans the core and inner wing of the Ca II K-line, hence our Ca K images show emission in this strong spectral line, which is formed in the chromosphere and upper photosphere. Sample images appear in Preminger et al (2001).  All images exhibit bright and dark features on a "quiet" background. These features are manifestations of magnetic active regions at different heights in the solar atmosphere. Dark sunspots are the most prominent features on the red and blue continuum images, while both sunspots and bright plages are distinct in the Ca II K-line images.

## 2.2 Data analysis

SFO data reduction and analysis procedures are described in detail in Walton et al. (1998) and Preminger et al. (2002). Presented here is a brief overview.  SFO images are suited to relative photometry:  for each image, a quiet-Sun limb-darkening curve is computed, then used to produce a contrast map. For image pixel i, we define its contrast, $C_i$, as:

$$C_i \equiv \frac{I_i}{I(\mu_i)} - 1, \qquad (1)$$

where $I_i$ is the observed intensity of pixel i, $\mu_i$ is the cosine of the heliocentric angle at pixel i, and  $I(\mu)$ is the quiet-Sun limb darkening curve for that image. On the contrast map, quiet Sun pixels have a contrast of zero, and solar features that are bright or dark relative to the quiet Sun have positive and negative contrasts respectively. The photometric sum Σ is given by:

$$\Sigma = \sum_i C_i \phi(\mu_i), \qquad (2)$$

where the sum extends over *all* image pixels, $C_i$ is the contrast of pixel i, and $\phi(\mu_i)$ is the limb darkening curve for that image, normalized to unit integral over the hemisphere. Thus for each image, Σ is the relative change (in parts per million) in the disk-integrated intensity of the image, due to the presence of solar features. We measure Σ for all of our images, and we use subscripts r, b and K to refer to the measurement on continuum red, continuum blue and Ca II K-line images, respectively. We note that the photometric sum includes the contributions of all solar features, bright and dark; noise pixels cancel in the summation, thus the photometric sum is sensitive to low contrast features that are difficult to isolate with threshold techniques. However, the accuracy of the photometric sum does depend on the correct identification of the quiet-Sun. Preminger et al (2002) show that a slight bias of the quiet-Sun is present in the contrast maps, due to the presence of numerous low-contrast bright features at solar maximum, but the effect of this bias can be assessed and corrected for. We revisit the issue of bias in Sections 2.3 and 3 of this manuscript.

## 2.3 SFO Results

In what follows we use the term 'brightness' to mean disk-integrated intensity. The photometric sum measures the change in Sun's brightness, due to the presence of magnetic features, relative to the brightness the solar disk would have if no features were present. $\Sigma_r$ and $\Sigma_b$ measure the relative change, due to active regions, in visible continuum brightness, while $\Sigma_K$ measures the relative change, due to active regions, in the Ca II K-line brightness.

Figure 2 shows the photometric sums $\Sigma_r$, $\Sigma_b$ and $\Sigma_K$, computed from SFO images. All three sums show large short-term variations that correspond to rotational modulation by bright and dark active regions, but short-term variability is most pronounced in the visible continuum, where sunspots cause very large dips in brightness as they transit the solar disk. In other studies we examined the short-term effects of active regions on radiative flux (Preminger and Walton 2005; 2006; Preminger et al. 2010). Here, we concentrate on the average effects of solar active regions over timescales of months to years. In Figure 2, the 81-day running mean is superimposed on the data, to highlight long-term trends.

As Figure 2 shows, $\Sigma_K$ is highest at solar maximum, i.e. the average effect of solar active regions is to enhance Ca II K-line brightness. We conclude from this plot and from Figure 1 that the Sun's relative brightness in the Ca II K-line is directly correlated with solar activity and TSI over time-scales of months to years. This is expected: magnetic activity has been shown to reduce chromospheric spectral-line blanketing, thus Ca II K-line brightness is directly correlated with the average photospheric magnetic flux density, which is greatest at solar maximum (Schrijver and Harvey 1989; Harvey and White 1999; Ortiz and Rast 2005; Preminger et al. 2010).

In Figure 2, $\Sigma_r$ shows the effect of solar activity on the Sun's red continuum brightness. Contrary to $\Sigma_K$, $\Sigma_r$ shows no enhancement at all at solar maximum: it seems almost flat. However, the running mean of $\Sigma_r$ is slightly anti-correlated with solar activity. We conclude that time-averaged red continuum brightness is diminished at solar maximum. The pattern of $\Sigma_b$ is similar to that of $\Sigma_r$: blue continuum variability is also anti-correlated with solar activity, on timescales of months to years. We infer from the red and blue photometric sums that the Sun's visible continuum brightness is slightly anti-correlated with solar activity, over solar-cycle timescales.

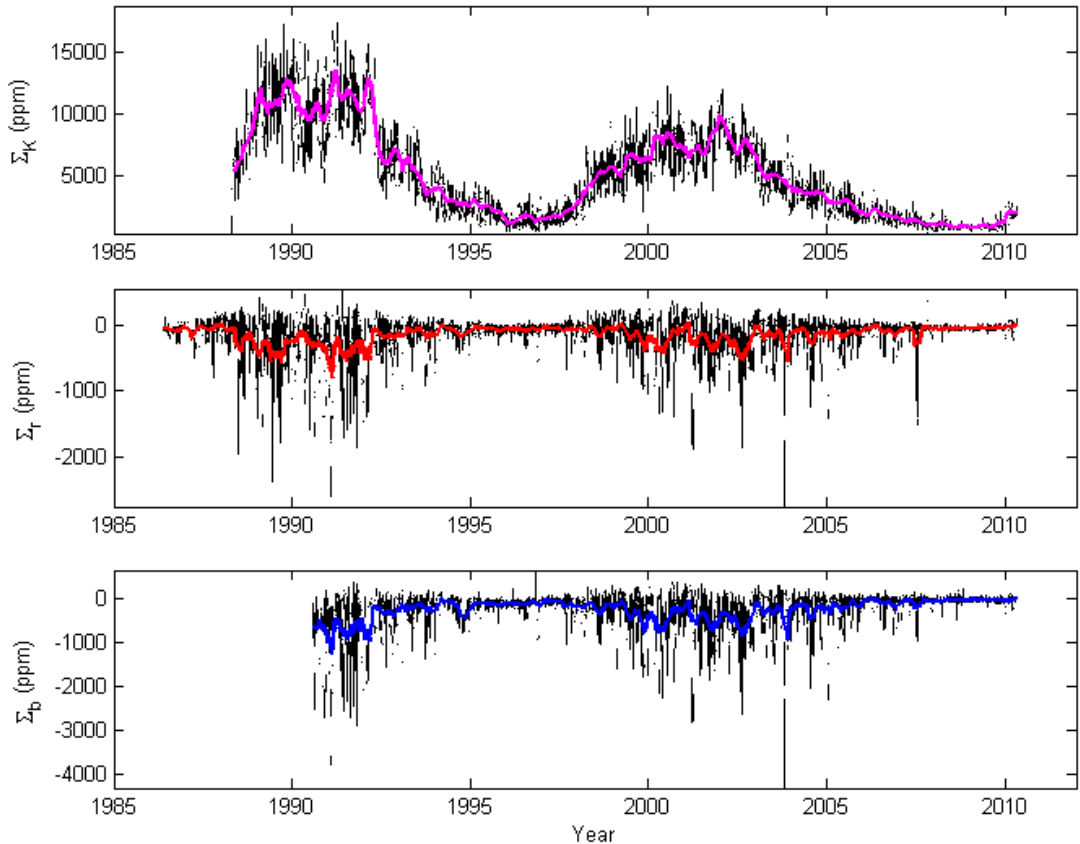

Figure 2. Top: The Ca II K-line photometric sum, $\Sigma_K$, from Ca II K-line images. Middle: The red photometric sum, $\Sigma_r$, from red continuum images. Bottom: the blue photometric sum, $\Sigma_b$, from blue continuum images. $\Sigma$ measures the relative change (in parts per million) in the disk-integrated intensity of the Sun due to the presence of solar features. The bold curves are the 81-day running means of the data.

We now address the issue of bias in the determination of the photometric sums. As stated, an accurate measurement of $\Sigma$ requires accurate identification of the quiet Sun. When the level of solar activity is high, there are small, low-contrast bright features present on the disk, tending to bias the intensity level determined for the quiet-Sun. In Preminger et al. (2002) we assessed the effect of this bias: it causes a small, positive, cycle-dependent trend in the data to be removed. To correct for this bias, we must reduce the long-term trend of $\Sigma_r$ by ~50%, and increase the long-term trend of $\Sigma_K$ by ~7%. This correction does not affect our qualitative assessment of the spectral variability: $\Sigma_K$ remains almost the same, while $\Sigma_r$ is still negatively correlated with the solar cycle, albeit less so.

## 3. Brightness vs. activity for the Sun

Studies show it is reasonable to conclude that solar magnetic activity accounts for most, if not all, solar radiative variability on time-scales of days to decades, while the 'quiet-Sun' background remains constant (Domingo et al. 2009 have an extensive review; Livingston et al. 2005). Now assuming this is the case, the photometric sum measures *all* variability in a given spectral band, and we can use this quantity to evaluate activity-brightness correlations for the Sun-as-a-star.

From Figures 1 and 2, the relationship between solar brightness and activity is clearly complex. The activity-related brightness variations in different parts of the spectrum are not similar to each other, and it is not obvious how changes in specific brightness are related to changes in bolometric brightness. However, we can use the photometric quantities, $\Sigma_r$ and $\Sigma_K$, to model changes in total solar irradiance ($\Delta$TSI) according to:

$$\Delta TSI(t) = a\Sigma_r(t) + b\Sigma_K(t) \tag{3}$$

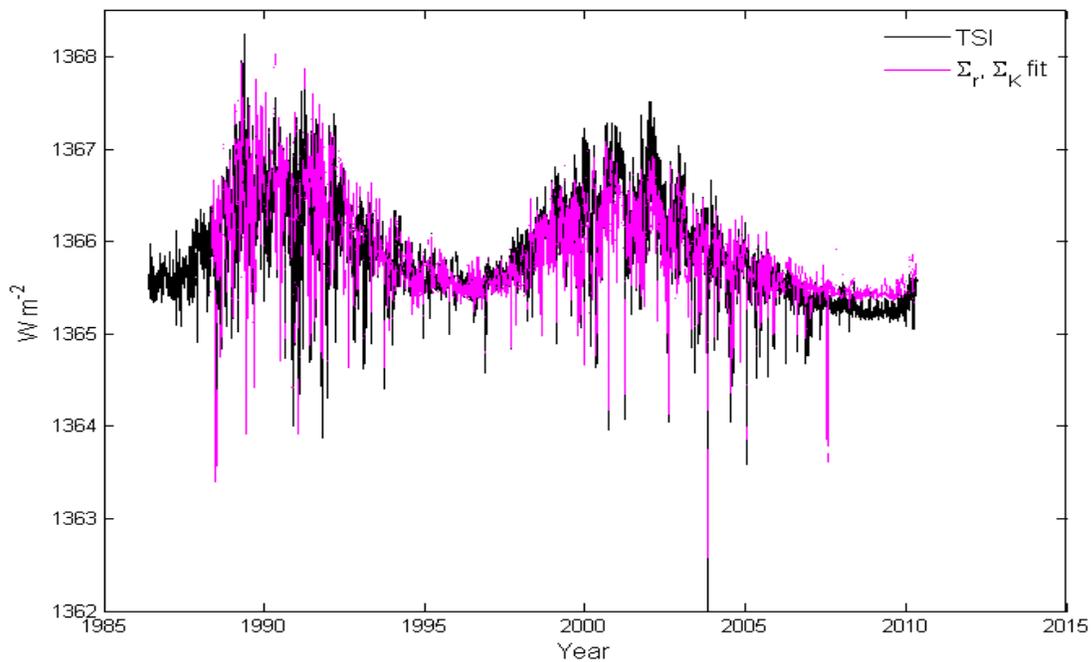

Figure 3. TSI and model fit using SFO data, $R^2$ = 0.84 (1988 – 2010)

This type of model was first used by Preminger et al. (2002) to reconstruct TSI variations during cycle 22. Figure 3 compares this simple model to TSI variations from 1988 – 2010. The fit is very good: the model, based on ~8700 data points, has a regression coefficient $R^2$ = 0.85. This is comparable to the goodness-of-fit achieved by other empirical and semi-empirical models of TSI (Domingo et al. 2009). We interpret the physical significance of Equation 3 as follows: to a first approximation, the net change in the Sun's bolometric flux is the linear sum of 2 components: the change in photospheric continuum flux (represented by $a\Sigma_r$ and the change spectral line-blanketing from the chromosphere and upper photosphere (represented by $b\Sigma_K$). All changes are induced by magnetic active regions. Equation 3 is a rather simplified model, but it can reconstruct 85% of TSI variability over a period of 2 solar cycles, which indicates that it incorporates the primary components of the Sun's irradiance variability in terms of energy.

Figure 4 shows the relative variability of ΔTSI(t) and the two contributing components from Equation 3. The pronounced short-term dips in TSI correspond to negative changes in the continuum component, $a\Sigma_r(t)$, that occur whenever a sunspot group transits the solar disk. But the continuum component is clearly not the primary source of longer-term TSI variations. The long-term variability is in the spectral lines, represented by $b\Sigma_K(t)$. This aspect of solar spectral variability was also pointed out in Preminger et al. (2002).

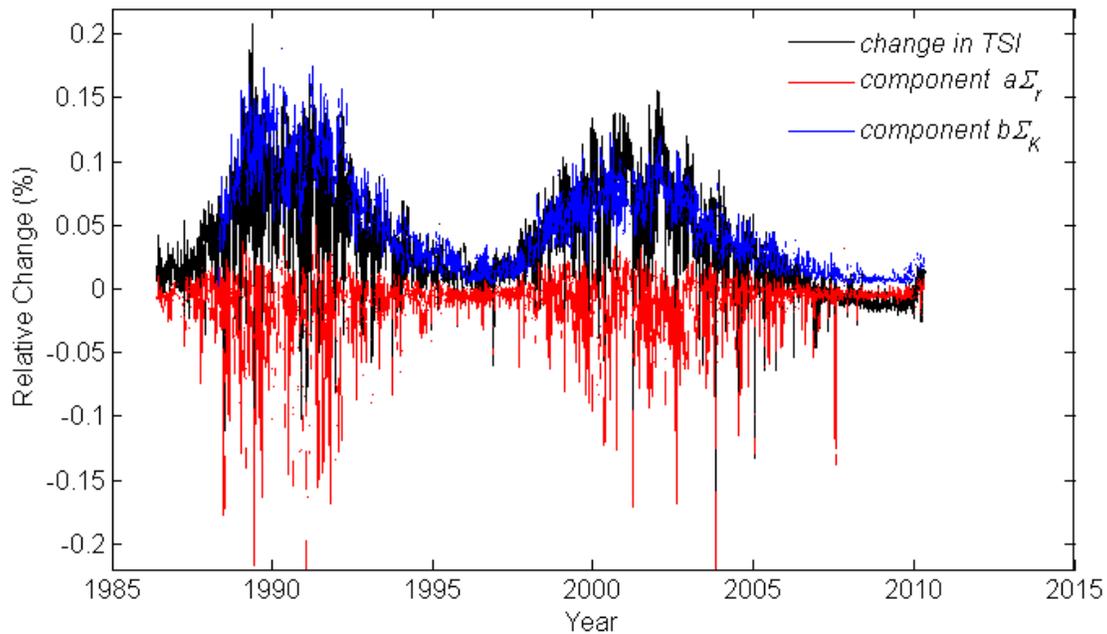

Figure 4. The relative change in TSI and the 2 primary contributing components, computed from Equation 3. We infer that the component $a\Sigma_r$ represents the change in continuum irradiance and the component $b\Sigma_K$ represents the change in spectral lines.

Long-term activity-brightness correlations become clear if we plot the time-average of the different components. For our best quantitative assessment, we first correct the photometric sums for quiet-Sun bias, as discussed in Section 2. Then we re-compute the components of the fit from Equation 3. Figure 5 shows the 81-day running mean of ΔTSI and the new fit components. From the figure we conclude that the bolometric and spectral line brightness are strongly enhanced by solar activity, on the order of 0.1% at solar maximum, while continuum brightness is slightly diminished, on the order of –0.02%.

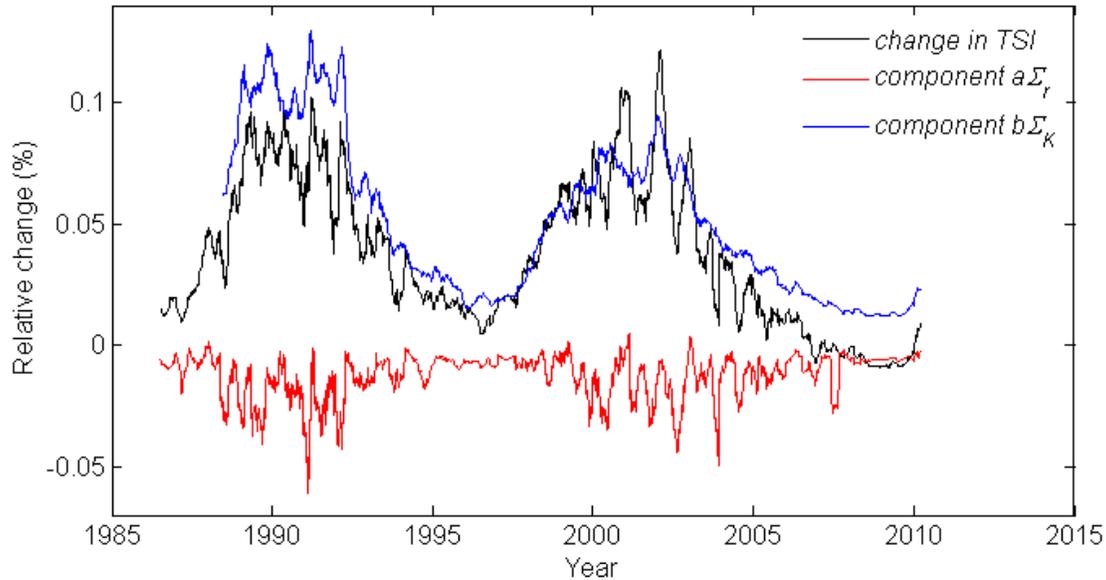

Figure 5. 81-day running mean of the relative change in TSI and the 2 primary contributing components. We infer that $a\Sigma_r$ represents the change in continuum irradiance and $b\Sigma_K$ represents the change in spectral lines. Here, the components have been corrected for the effects of quiet-Sun bias.

## 4. The Sun vs. Solar analogs

Many Sun-like stars appear to have magnetic activity cycles. Generally, observations of Ca HK emission are used to infer the level of stellar magnetic activity, and stellar brightness is measured using the Strömgren b and y photometric system (Hall et al. 2009; Radick et al. 1998; Lockwood et al. 2007). Observations of Sun-like stars are carried out during an observing season, and the seasonal mean is studied. An observing season lasts about three months, therefore we can compare the results for stellar variability with the 81-day average variability of the Sun.

Observations show that solar analogs appear to fall into two groups: stars with high activity levels (i.e. large variation in Ca II HK emission) tend to have an inverse correlation between b+y brightness and HK emission, while less active stars tend to have a positive correlation. Let us compare these stellar observations to those of the Sun.

The connection between magnetic active regions and enhanced emission in chromospheric lines such as the Ca II H and K lines is already well-established (e.g. Leighton 1959; Schrijver et al. 1989). Disk-integrated Ca K-line variability as quantified by $\Sigma_K$ can be considered to be directly related to the Ca HK brightness index commonly used to measure chromospheric activity for stars. Both quantities are good measures of magnetic activity.

The following compares filters used for stellar Strömgren b and y photometry with the blue and red filters used at SFO for solar photometry:

| Filter | Strömgren b | Strömgren y | SFO blue | SFO red |
|---|---|---|---|---|
| Wavelength (nm) | 467 | 547 | 472 | 672 |
| Bandpass (nm) | 18 | 23 | 10 | 10 |

For the Strömgren filters, photospheric spectral lines block ~8% and ~11% of the continuum flux in the passbands of the y and b filters, respectively (Conti & Deutsch, 1966). Thus SFO blue and red filters are comparable to the Strömgren b and y filters: all measure continuum brightness because they are broad and located above the point (about 450 nm) where line-blanketing becomes important (Crawford 1987). Stellar b+y brightness is a measure of visible continuum brightness.

The most active Sun-like stars, those observed to have an inverse correlation between b+y brightness and HK emission on timescales of months to years, display an inverse relationship between visible continuum and spectral line flux over these timescales. This is the same pattern we see for the Sun! Thus, if solar analogs fall into 2 groups, we must conclude that the Sun belongs with the *more* active group.

This conclusion is opposite to that reached by Lockwood et al. (2007). The problem is that, for Sun-like stars, bolometric brightness is inferred, not measured. It is assumed that the star radiates like a black body, and variations in flux arise from changes in its effective temperature:

$$\Delta\left(\frac{b+y}{2}\right)(mag) = \frac{\Delta F_{bol}}{F_{bol}} \tag{4}$$

For active Sun-like stars, b+y flux is low when activity is high. Applying Equation 4 leads to the conclusion that the bolometric brightness of these stars is reduced by activity. However, our analysis of solar variability shows that Equation 4 is invalid for the Sun, and likely also invalid for Sun-like stars.

## 5. Summary and Conclusions

The photometric quantity $\Sigma$ measures the effect of all solar features (bright and dark) on the relative brightness of the solar disk. We examine $\Sigma$ for two solar-cycles' worth of red, blue, and Ca II K-line photometric images from the San Fernando Observatory. $\Sigma_r$ and $\Sigma_b$ measure visible continuum variability, due to solar activity, while $\Sigma_K$ measures variability in the Ca II K spectral line, caused by solar activity. When we average over timescales of a few months, we find that the net effect of solar features is to enhance the brightness of the solar disk in the Ca II K-line and to diminish the brightness of the disk in the visible continuum spectral range. We originally reported this result based on SFO data for solar cycle 22 (Preminger et al. 2002); our data for solar cycle 23 now confirm the result. New results from the SIM experiment on the SORCE satellite appear to corroborate our findings (Harder et al. 2009).

We construct a model of the total solar irradiance, TSI, of the form $\Delta TSI = a\Sigma_r + b\Sigma_K$. This model successfully reproduces 85% of all TSI variability over the time period 1988 – 2010. Physically, the model implies that two primary components of bolometric variability exist: variations in continuum, photospheric irradiance (represented by $a\Sigma_r$) and variations in spectral lines (represented by $b\Sigma_K$). Over solar-cycle timescales, the Sun's bolometric irradiance is directly correlated with spectral line irradiance, but inversely correlated with continuum irradiance. The quiet-Sun irradiance is assumed constant.

We compare the activity-brightness relations determined for the Sun to those observed for Sun-like stars. For those stars, activity is measured using the Ca II HK index while brightness is measured using Strömgren b+y photometry. These 2 measures are anti-correlated for very active stars (those with large periodic changes in the Ca HK index), and positively correlated for less active stars. The b+y brightness was assumed to be directly related to bolometric brightness, making the behavior of more active stars appear to be the opposite of the Sun's. But, b+y brightness is actually a measure of visible continuum brightness. Thus Sun-like stars for which b+y brightness is inversely correlated with activity are behaving in a very Sun-like way.

If solar analogs are indeed Sun-like, we might expect the magnetic active regions to be similar in nature, and to affect the radiative properties of the stellar atmosphere in a similar way. And, indeed, the relationship between photospheric magnetic flux and radiative flux from the chromosphere or corona is similar for the Sun and solar analogs (Schrijver et al. 1989; Pevstov et al. 2003). Therefore, we should not be surprised to find Sun-like stars for which activity and photospheric radiative flux are anti-correlated, as they are for the Sun. How, then, can we explain that some Sun-like stars exhibit activity directly correlated with photospheric brightness? It could be due to the angle from which we view these stars: we know that solar active regions appear at mid-to-low latitudes only, and that sunspots decay into facular regions which appear bright in continuum images when they are on the solar limb. In a polar view of the Sun, the active regions would circle the limb; the average brightness of the disk might then show small variations, in phase with activity, even for visible continuum wavelengths. The stars for which activity and brightness are directly correlated are those with small HK variability, making this explanation plausible.

The results of this study may help unite observations of activity vs. brightness for the Sun and Sun-like stars. In doing so, they may help answer a persistent question regarding the source of the Sun's bolometric irradiance variations: is the enhancement at solar maximum really caused by the magnetic features discernable on Ca II K-line (chromospheric) images? Models such as that described by Equation 3, which assume that known magnetic features are responsible for all irradiance variability, correlate very well with observations, but correlation does not prove causation. Could the real source of long-term variability be small-scale, low-contrast photospheric features that we cannot resolve on our continuum images, or could the solar cycle change in TSI be driven by co-temporal changes in the effective temperature or radius of the photosphere? We propose that if the mechanism of irradiance variability is the same for the Sun and Sun-like stars, we can rule out these possibilities: for, if Sun-like stars are indeed like the Sun, their bolometric brightness should increase with activity, on stellar-cycle timescales. This increase cannot have its origin in the photosphere, since total visible continuum brightness (disk-integrated, from *all* sources) is observed to decrease at the same time. Therefore long-term bolometric variability must be caused by the effects of magnetic active regions on spectral line emission from the higher layers of the stellar atmosphere. In order to confirm this hypothesis, it will be necessary to measure the bolometric variability of Sun-like stars.

## Acknowledgments

This research has been partially supported by NASA Living with a Star grant NNX07AT19G, NASA grant NNX11AB51G and by NSF grant ATM-0848518.